\documentclass{article}

\usepackage{natbib}
 \bibpunct[, ]{(}{)}{,}{a}{}{,}%

\usepackage{url}
\usepackage{setspace}

\usepackage[top=1in, bottom=1in, left=1in, right=1in]{geometry}

\title{Data Analytics in Operations Management: A Review}
\author{Velibor V. Mi\v{s}i\'{c} \thanks{Anderson School of Management, University of California, Los Angeles, 110 Westwood Plaza, Los Angeles, CA, 90095; \texttt{velibor.misic@anderson.ucla.edu} }  \and Georgia Perakis \thanks{Sloan School of Management, Massachusetts Institute of Technology, 77 Massachusetts Avenue, Cambridge, MA, 02139; \texttt{georgiap@mit.edu}} }
\date{March 29, 2019}

\begin{document}

\maketitle

\begin{abstract}
Research in operations management has traditionally focused on models for understanding, mostly at a strategic level, how firms should operate. Spurred by the growing availability of data and recent advances in machine learning and optimization methodologies, there has been an increasing application of data analytics to problems in operations management. In this paper, we review recent applications of data analytics to operations management, in three major areas -- supply chain management, revenue management and healthcare operations -- and highlight some exciting directions for the future. 
\end{abstract}

\section{Introduction}

Historically, research in operations management has focused on models. These models, based on microeconomic theory, game theory, optimization and stochastic models, have been mostly used to generate strategic insights about how firms should operate. One of the reasons for the prevalence of this model-based approach in the past has been the relative scarcity of data, coupled with limitations in computing power. 

Today, there has been a shift in research in operations management. This shift has been primarily driven by the increasing availability of data. In various domains -- such as retail, healthcare and many more-- richer data is becoming available that is more voluminous and more granular than ever before.
At the same time, the increasing abundance of data has been accompanied by methodological advances in a number of fields. The field of machine learning, which exists at the intersection of computer science and statistics, has created methods that allow one to obtain high quality predictive models for high-dimensional data, such as L1 regularized regression (also known as LASSO regression; \citealt{tibshirani1996regression}) and random forests \citep{breiman2001random}. The field of optimization has similarly advanced: numerous scientific innovations in optimization modeling, such as robust optimization \citep{bertsimas2011theory}, inverse optimization \citep{ahuja2001inverse} and improved integer optimization formulation techniques \citep{vielma2015mixed} have extended both the scope of what optimization can be applied to and the scale at which it can be applied. In both machine learning and optimization, researchers have benefited from the availability of high quality software for estimating machine learning models and for solving large-scale linear, conic and mixed-integer optimization problems.

These advances have led to the development of the burgeoning field of \emph{data analytics} (or \emph{analytics} for short). The field of analytics can most concisely be described as using \emph{data} to create \emph{models} leading to \emph{decisions} that create \emph{value}. In this paper, in honor of the 20th anniversary of \emph{Manufacturing \& Services Operations Management}, we highlight recent work that applies the analytics approach to problems in operations management. We divide our review along three major application areas: supply chain management (Section~\ref{sec:scm}), where we cover location, omni-channel and inventory decisions; revenue management (Section~\ref{sec:rm}), where we cover choice modeling and assortment optimization, pricing and promotion planning, and personalized revenue management; and healthcare (Section~\ref{sec:healthcare}), where we cover applications at the policy, hospital and patient levels. We conclude in Section~\ref{sec:conclusion} with a discussion of some future directions -- such as causal inference, interpretability and ``small data'' methods -- that we believe will be increasingly important in the future of analytics in operations management. 

Due to the page limitations imposed by this anniversary issue, our review is necessarily brief and only covers a small set of representative examples in each application area. As a result, there are many other excellent papers that apply analytics to OM that we were unable to include in this review; we apologize in advance to those authors whose papers we have not cited.

\section{Analytics in Supply Chain Management}
\label{sec:scm}

Many major problems in supply chain management are being re-examined under the lens of analytics; in this subsection, we focus on location and omnichannel decisions (Section~\ref{subsec:scm_location}) as well as inventory decisions (Section~\ref{subsec:scm_inventory}). 

\subsection{Location and Omnichannel Operations}
\label{subsec:scm_location}

In this subsection we discuss the application of analytics to omnichannel and more generally, location problems. The term ``omnichannel'' refers to the integration of an e-commerce channel and a network of brick-and-mortar stores. This integration allows a retailer to do cross-channel fulfillment, that is, fulfill online orders in any store location, leading to interesting location and inventory management problems. \cite{glaeser2018optimal} considers a location problem faced by a real ``buy online, pickup in store'' retailer, that fulfills online orders through delivery trucks parked at easily accessible locations (e.g., at schools or parking lots). The retailer needs to decide at which locations and times to position its trucks, in order to maximize profit. To solve this problem, the paper first builds a random forest model \citep{breiman2001random} to predict demand at a given location at a given time, using a diverse set of independent variables such as demographic attributes of the location (total population, population with a post-secondary degree, median income, and so on), the retailer's operational attributes (such as whether or not the retailer offers home delivery in this location) and other location attributes (e.g., number of nearby competing businesses). Then, the paper uses fixed effects regression to account for cannibalization effects. Using the retailer's data, the paper shows how a heuristic approach based on greedy construction and interchange ideas together with the combined random forest and fixed effects model, leads to improvements in revenue of up to 36\%. 

\cite{acimovic2014making} study how to manage the omnichannel operations of a large online retailer. They address the fundamental questions of what is the best way to fulfill each customer's order in order to minimize average outbound shipping costs. The paper develops a heuristic that makes fulfillment decisions by minimizing the immediate outbound shipping cost plus an estimate of future expected outbound shipping costs. These estimates are derived from the dual values of a transportation linear program. Through experiments on industry data, their model captures 36\% of the opportunity gap assuming clairvoyance, leading to reductions in outbound shipping costs on the order of 1\%. 
In the same spirit, \cite{avrahami2014matching} discuss their collaboration with the distribution organization within the Yedioth Group in Israel, to improve how Yedioth distributes print magazines and newspapers. This paper uses real-time information on the newspaper sales at the different retail outlets to enable pooling of inventory in the distribution network. The methods developed were implemented at Yedioth and generated substantial cost savings due to both a reduction in magazine production levels and a reduction in the return levels.

In a different domain, \cite{he2017service} consider the problem of how to design service regions for an electric vehicle sharing service. A number of car sharing firms, such as Car2Go, DriveNow and Autolib, offer electric vehicles to customers. Providing such a service in an urban area requires the firm to decide the regions in which the service will be offered, i.e., in which regions of the urban area are customers allowed to pick up and drop off a vehicle; this, in turn, informs major investment decisions (how large the fleet needs to be and where charging stations should be located). The main challenge in this setting comes from the fact that there is uncertainty in customer adoption, which depends on which regions are covered by the service. To solve this problem, the paper formulates an integer programming problem where customer adoption is represented through a utility model; due to the limited data from which such a utility model can be calibrated, the paper uses distributionally robust optimization \citep[see, e.g.,][]{delage2010distributionally} to account for the uncertainty in customer adoption. Using data from Car2Go, the paper applies the approach to designing the service region in San Diego, and shows how the approach leads to higher revenues than several simpler, managerially-motivated heuristic approaches to service region design.

\subsection{Inventory management}
\label{subsec:scm_inventory}

\cite{ban2018big} consider a data-driven approach to inventory management. The context that the paper studies is when one has observations of demand, together with features that may be predictive of demand, such as weather forecasts or economic indicators like the consumer price index. To make inventory decisions in this setting, one might consider building a demand distribution that is feature-dependent, and then finding the optimal order quantity for the distribution corresponding to a given realization of the features. Instead, the paper of \cite{ban2018big} proposes two alternate approaches. The first, based on empirical risk minimization, involves finding the order quantity by solving a single problem, where the decision variables is the decision rule that maps the features to an order quantity, and the objective is to minimize a sample-based estimate of the cost. The second approach involves using kernel regression to model the conditional demand distribution, and applying a sorting algorithm to determine the optimal order quantity. The paper derives bounds on the out-of-sample performance of both approaches, and applies the approaches to the problem of nurse staffing in a hospital emergency room using data from a large teaching hospital in the United Kingdom. The paper finds that the proposed approaches outperform the best-practice benchmark by up to 24\% in out-of-sample cost.

In a similar vein, \cite{bertsimas2018predictive} develop a general framework for solving a collection of sample-average approximation (SAA) problems, where each SAA problem has some contextual information. For example, in an inventory setting, a firm might be selling different products, which are described by vectors of attributes, for which the firm has historical observations of demand; the firm would then like to determine an order quantity for a (potentially new) product. A naive approach to this problem would be to simply pool all of the data together and ignore the contextual information. Instead, \cite{bertsimas2018predictive} propose building a machine learning model first to predict the uncertain quantity as a function of contextual information; in our inventory example, this would amount to predicting demand as a function of the product attributes. Then, instead of solving the SAA problem using all of the data on the uncertain parameter, one uses the machine learning model to obtain a context-specific conditional distribution, and then solves the SAA problem with respect to this conditional distribution. In the inventory setting, consider a CART tree \citep{breiman1984classification} for predicting demand for a given product attribute vector. If one runs the product's attribute vector down the CART tree, one will obtain a point prediction, but by considering the historical demand observations contained in that leaf, one also obtains an estimate of the conditional distribution. The paper develops theoretical guarantees for this general procedure for specific types of machine learning models, proving asymptotic optimality of the procedure as the number of observations grows. The paper also applies the approach to a real distribution problem for a media company that needs to manage inventories of different products (specifically, movie CDs, DVDs and Bluray discs) across different retail locations. The paper builds machine learning models of demand as a function of the store location, properties of the movie (such as its genre, rating on Rotten Tomatoes, box office revenue and so on) and other large-scale information (such as localized Google search queries for the movie). The paper shows how the approach is able to close the cost gap between the naive approach (which does not consider contextual data) and the perfect foresight approach (which knows demands before they are realized) by 88\%. 

\cite{ban2018dynamic} consider the problem of dynamic procurement. In this problem, one has to decide how much of a product to order over a finite time horizon from different sources which have different lead times and different costs, so as to meet uncertain demand over the horizon. The problem setting studied in the paper is motivated by a practical problem faced by Zara, one of the world's largest fast-fashion retailers. The paper propose an approach that involves using linear and regularized linear regression to predict the demand for a given product, using the historical trajectories of other products. Then, the paper formulates a multi-stage stochastic programming problem, where the scenario tree is generated using the previously-estimated predictive model. Using real data from Zara, the paper shows how the existing approach, which ignores covariate information, can lead to costs that are 6-15\% higher than the proposed approach.

At the core of many inventory problems -- and operations management problems in general -- lies the problem of demand estimation. The results of this estimation  can then  be used, among others, in order to address pricing, distribution, and inventory management questions.  \cite{baardman2017leveraging} develop a data driven approach for predicting demand for new products. The paper develops a machine learning algorithm that identifies comparable products in order to predict demand for the new product. The paper demonstrates both analytically and empirically, through a collaboration with Johnson \& Johnson, the benefits of this approach.

\section{Analytics in Revenue Management}
\label{sec:rm}

Analytics is having a major impact on how firms make decisions about how to sell their products to customers; in this section, we discuss applications in choice modeling and assortment optimization (Section~\ref{subsec:rm_choice}), pricing and promotion planning (Section~\ref{subsec:rm_pricing}) and personalized revenue management (Section~\ref{subsec:rm_personalized}).

\subsection{Choice modeling and assortment optimization}

\label{subsec:rm_choice}

Assortment optimization refers to the problem of deciding which products to offer to a set of customers, so as to maximize the revenue earned when customers make purchase decisions from the offered products. A major complicating factor in assortment optimization is the presence of substitution behavior, where customers may arrive with a particular product in mind, but purchase a different product if that first product is not available. %

The paper of \cite{farias2013nonparametric} proposed a nonparametric model for representing customer choice behavior, by modeling the customer population as a probability distribution over rankings of the products. Such a model is able to represent any choice model based on random utility maximization, which includes the majority of popular choice models (e.g., the MNL model, the nested logit model and the latent-class MNL model). The paper shows, using data from a major auto manufacturer, how the proposed approach can provide more accurate predictions than traditional parametric models such as the MNL model. %
In later work, a number of papers have considered how to efficiently find assortments that maximize expected revenue under such a model. From a tractability standpoint, \cite{aouad2018approximability} showed that the assortment problem is NP-Hard even to approximate (APX-Hard) and developed an approximation algorithm; \cite{feldman2018assortment} later provided an approximation algorithm with an improved guarantee. For specific classes of ranking-based models, \cite{aouad2015assortment} developed efficient approaches based on dynamic programming for solving the assortment optimization problem exactly. In a different direction, the paper of \cite{bertsimas2018exact} proposed a new mixed-integer optimization formulation for the problem as well as a specialized solution approach, based on Benders decomposition, for solving large-scale instances efficiently. Using a product line design problem based on a real conjoint analysis data set, the paper shows how a large-scale product line design problem that required one week of computation time to solve in earlier work is readily solved by the proposed approach in ten minutes.

In addition to the ranking-based model, the paper of \cite{blanchetgallegogoyal} recently introduced a Markov chain-based choice model. In this model, substitution from one product to another is modeled as a state transition in a Markov chain. The paper shows that under some reasonable assumptions, the choice probabilities computed by the Markov chain based model are a good approximation of the true choice probabilities for any random utility-based choice model, and are exact if the underlying model is a generalized attraction model (GAM) of which the MNL model is a special case.  In addition to the theoretical bounds, the authors conduct numerical experiments and observe that the average maximum relative error of the choice probabilities of their model with respect to the true probabilities for any offer set is less than 3\% where the average is taken over different offer sets.

\subsection{Pricing and promotion planning}
\label{subsec:rm_pricing}

An exciting example of how analytics is being applied in pricing is the paper of \cite{ferreira2015analytics}, which studies the pricing of Rue La La's weekly sales. Rue La La is an online fashion retailer that sells designer apparel items (called \emph{styles}) at significantly discounted prices in weekly, limited-time sales (called \emph{events}). The paper develops a strategy for optimally pricing styles in an event that combines machine learning and optimization. The machine learning component of the approach involves developing a bagged regression tree model \citep{breiman1996bagging} for predicting the demand of a style based on the price of the style and the average price of other styles in the event. The optimization component involves formulating an integer optimization problem where the goal is to find the prices of the styles that maximize the total revenue from all styles, which is just the product of the price of the style and the demand predicted by the bagged regression tree model. This optimization problem is, in general, very difficult. However, if one fixes the average price of all styles to a given value, then the problem simplifies, and one can solve a small number of smaller integer optimization problems to find the optimal collection of prices. The paper reports on a live pilot experiment to test the methodology at Rue La La, which resulted in an increase in revenues of about 10\%.

Another example where analytics has made a difference in the field of pricing relates to promotion planning.  \cite{cohen2017impact}, \cite{cohen2017optimizing} as well as  \cite{hillelcohenetal2019a},  \cite{hillelcohenetal2019b}   consider how analytics applies to price promotion planning first for a single item and subsequently, for multiple items. Price promotion planning is concerned with when and how deeply to promote each item over a time horizon. The problem is challenging because sales of an item in a given time period are affected by not only the price of that item in that period, but also what the prices of that item were before then, and what the prices of other items are. For example, a promotion on a coffee SKU (Stock Keeping Unit) today may result in a large increase in sales if it has not been recently promoted, but may only result in a modest change if the coffee SKU has been promoted heavily in the preceding weeks (as customers may, for example, have stockpiled the item during earlier promotion periods). The paper of \cite{cohen2017impact} considers the single item problem and introduces a general mixed-integer nonlinear optimization formulation of the problem, where one first learns a demand model from available data, and then uses that demand model in the objective of the optimization model. The objective in this body of work corresponds to profit and the constraints reflect common business rules imposed by retailers in practice (for example, limits on the number of successive promotions over any fixed window). While the optimization problem is in general theoretically intractable, \cite{cohen2017impact}  show how an approximate formulation, based on linearizing the nonlinear objective, leads to an efficiently solvable mixed-integer linear problem. The paper shows how the suboptimality gap of the promotion price schedule derived from the approximation can be bounded for different classes of demand functions. This approach was also later extended to the multi-item price promotion planning problem in \cite{cohen2017optimizing}.

More recently, \cite{hillelcohenetal2019a} and \cite{hillelcohenetal2019b} have considered a demand model that only considers a limited set of pricing features; in particular, one considers the price of the item in the current time period, the price in the previous period as well as the minimum price within a bounded set of past periods. These papers consider the pricing problem under this model, which is referred to as the \emph{bounded memory peak-end} demand model. These papers show how one can exploit the structure arising from these price features to introduce a compact dynamic programming formulation of the pricing problem, allowing the problem to be efficiently solved. In the multi-item version of the problem, the papers show that the problem is NP-Hard in the strong sense, but establish a polynomial time approximation scheme (PTAS).
Furthermore,  \cite{hillelcohenetal2019b} show that even if the true underlying demand model does not follow the bounded memory peak-end demand model, but rather follows some of the more traditional models in the literature, the bounded memory peak-end demand model still performs very well; the paper provides a theoretical bound for this result, which is tested with data. Using data from a grocery retailer, these papers show how the optimization approaches introduced can lead to profit increases between 3-10\%. 

The previous papers deal with price promotions, which constitute one type of promotion. Retailers have access to many other methods of promotion (termed \emph{promotion vehicles}), such as flyers, coupons and announcements on social media websites (e.g., Facebook). The paper of \cite{baardman2018scheduling} considers how one should schedule such promotion vehicles so as to maximize profit. The paper shows that the problem is APX-Hard and subsequently introduces a nonlinear integer optimization formulation for scheduling promotion vehicles under various business constraints (such as limiting the promotion vehicles to use each time period, among others). The paper considers a greedy approach as well as a linear mixed-integer optimization approximation that is more accurate than the greedy method. Using real data, the paper illustrates the quality of the two approximation methods as well as the potential profit improvement, which is of the order of 2-9\%.

In the omnichannel retail setting, consumers can compare store prices with online prices, giving rise to challenging pricing problems. \cite{harshauichanco} study the price optimization aspect of omnichannel operations. Due to the presence of cross-channel interactions in demand and supply, the paper proposes two pricing policies that are based on the idea of ``partitions'' to the store inventory, which approximate how it will be used by the two different channels. These two policies are respectively based on solving either a deterministic or a robust mixed-integer optimization problem that incorporates several business rules as constraints. %
Through an analysis of a pilot implementation at large US retailer, the paper estimates that the methodology leads to a 13.7\% increase in clearance-period revenue.

\subsection{Personalized revenue management}
\label{subsec:rm_personalized}

Another major area of research is in personalized revenue management. Retailers have become increasingly interested in personalizing their products and services, including promotions. Personalization refers to the ability to tailor decisions to the level of individuals: in revenue management, this entails offering different prices, changing the product assortment, as well as other decisions such as promotions and product bundling. 

Many papers have recently considered the problem of personalized pricing.  %
\cite{chen2015statistical} consider a setting in which a firm sells a single product to a customer population, where each customer chooses to buy the product according to a logit model that depends on customer covariates and the price, and the firm has access to historical transaction records, where each record contains the covariates of the customer, the price at which the product was offered, and a binary indicator of whether the customer bought the product or not. The paper proposes a simple approach to personalized pricing: estimate the parameters of the logistic regression model using regularized maximum likelihood estimation, and then set the price for future customers so as to maximize the predicted expected revenue (price multiplied by purchase probability) under the estimated parameters. The paper bounds the suboptimality gap between the expected revenue of this policy and that of an oracle pricing policy that knows the customer choice model perfectly. Using airline sales data, the paper shows how the proposed method provides an improvement of over 3\% in revenue over a pricing policy that offers all customers the same price.

\cite{ban2017personalized} develop an algorithm for dynamic pricing of a single product, where arriving customers are described by a feature vector and the demand in each period is given by a model that depends on the price and only a subset of the features; the demand model allows for feature-dependent price sensitivity. In this setting, the paper proves a lower bound on the regret of any dynamic pricing algorithm in terms of the number of relevant features, and develops algorithms with near-optimal regret. The paper demonstrates, using data from an online auto loan company, how the proposed algorithms accrue more revenue than the company's current practice as well as other recently proposed dynamic pricing algorithms. 

Personalization of product offerings has also led to the need to develop new personalized demand models. Unfortunately, in practice (and especially in the retail space), the data available is not necessarily at the level of granularity needed to develop meaningful demand models at the personalized level. A potential path to developing personalized demand models is to use data on customers from social media, which can directly inform a retailer on how one customer's purchases affect the purchases of a different customer. However, for most retailers, acquiring this data is costly and poses challenges due to privacy concerns. \cite{baardman2018detecting} focus on the problem of detecting customer relationships from transactional data, and using them to offer targeted price promotions to the right customers at the right time. The paper  develops a novel demand model that distinguishes between a base purchase probability, capturing factors such as price and seasonality, and a customer trend probability, capturing customer-to-customer trend effects. The estimation procedure is based on regularized bounded variables least squares combined with the method of instrumental variables, a commonly used method in econometrics/causal inference \citep{angrist2008mostly}. The resulting customer trend estimates subsequently feed into a dynamic promotion targeting optimization problem, formulated as a nonlinear mixed-integer optimization problem. Although the problem is NP-Hard, the paper also proposes an adaptive greedy algorithm and proves that the customer-to-customer trend estimates are statistically consistent, but also that the adaptive greedy algorithm is provably good. Using actual fashion data from a client of Oracle, the paper shows that the demand model reduces the prediction error by 11\% on average, and the corresponding optimal policy increases profits by 3-11\%.

Besides the papers discussed above, there has been other recent research in the area of dynamic pricing with learning. \cite{javanmard2016dynamic} consider a single-product setting where in each period, there is only one customer that purchases the product if the price is below their valuation, and propose a regularized maximum likelihood policy that achieves near-optimal expected regret. \cite{cohen2016feature} consider a firm selling a sequence of arriving products, and each product is described by a feature vector. Each product is sold if its price is lower than the market value of the product, which is assumed to be linear in the features of the product. As more and more products arrive and the firm observes how the market responds to the product (buy/no buy decisions), the firm can refine a polyhedron representing the uncertainty in the parameters of the market's valuation model. The paper proposes a policy based on approximating this polyhedron with an ellipsoid and setting prices in a way that balances the revenue gained against the reduction in uncertainty, and show that this policy achieves a worst-case regret that scales favorably in the number of features and the time horizon.

There has also been much interest in personalized assortment planning. In addition to personalized pricing, \cite{chen2015statistical} develop similar theoretical guarantees as discussed above for the problem of assortment personalization based on a finite sample of data. \cite{golrezaei2014real} consider dynamic assortment personalization when there is a limited inventory of products. The difficulty in the problem arises from the limited inventory of products -- given a customer now, we may choose to offer them a particular product as part of an assortment, but if they choose to purchase that product, the inventory of that product will be depleted and we may run out of that product for later customers. The paper proposes a heuristic, called inventory balancing, which attains significantly improved revenues over simpler benchmarks. At the same time, there is also a growing literature that studies assortment personalization in a dynamic setting where one must learn the underlying choice model while making assortment decisions. For example, \cite{bernstein2018dynamic} consider a bandit approach to assortment personalization based on dynamic clustering. This approach assumes that there exists a clustering of customer ``profiles'' (vectors of attributes); however, the number of clusters and the clustering (mapping of profiles to clusters) are unknown. The approach involves modeling the distribution over clusterings using the Dirichlet Process Mixture model, which is updated in a Bayesian fashion with each transaction. 

Personalization has also been applied to other types of revenue management decisions. \cite{ettl2018data} consider the problem of making personalized price and product bundle recommendations over a finite time horizon to individual customers, so as to maximize revenue with limited inventory. The paper develops a number of approximation algorithms related to the inventory balancing heuristic \citep{golrezaei2014real} and Lagrangian relaxation methods for weakly coupled stochastic dynamic programs \citep{hawkins2003lagrangian}. Using two different data sets, one in airline ticket sales and one from an online retailer, the paper shows how the proposed approach can improve revenues over pricing and bundle strategies used in practice by 2-7\%. 

\section{Analytics in Healthcare Operations}
\label{sec:healthcare}

Machine learning and big data methodologies are beginning to also have a significant impact in healthcare operations. We divide our discussion in this part of the paper to problems at the policy level (Section~\ref{subsec:healthcare_policy}), the hospital level (Section~\ref{subsec:healthcare_hospital}) and the patient level (Section~\ref{subsec:healthcare_patient}).

\subsection{Policy-level problems}
\label{subsec:healthcare_policy}

A number of papers have considered the application of analytics to healthcare policy problems. The paper of \cite{aswani2018data} studies the Medicare Shared Savings Program (MSSP). The MSSP was introduced by Medicare to combat rising healthcare costs: Medicare providers that enroll in MSSP and that reduce their costs below a certain financial benchmark receive bonus payments from Medicare. However, few Medicare providers have been able to successfully reduce their costs in order to receive the bonus payments, due to the investment required. The paper of \cite{aswani2018data} models the interaction between Medicare and each provider as a  principal-agent problem, where Medicare sets the financial benchmark for each provider, and each provider decides what investment to make to reduce their costs. The authors use data from Medicare, as well as an estimation methodology based on inverse optimization \citep{ahuja2001inverse}, to learn the parameters of their principal-agent model. 
Using this estimated model, they propose an alternate contract, where the payment depends on both reducing cost below a financial benchmark and the amount invested, and show that this alternate contract can increase Medicare savings by up to 40\%.  

\cite{bertsimas2013fairness} consider the design of a dynamic allocation policy for deceased donor kidneys to waiting list patients. In the design of an allocation policy, a key objective is efficiency, i.e., the benefit garnered from allocating (transplanting) a kidney to a patient. However, another objective is fairness, i.e., making sure that patients with certain characteristics are not systematically underserved by the allocation policy. The paper proposes formulating the problem as a linear optimization problem, where the objective is to maximize efficiency (the aggregate match quality) subject to a fairness constraint, and uses linear regression to derive a scoring rule from the solution of this linear optimization problem to predict the fairness-adjusted quality of a match using attributes of both the patient and the donor kidney. The paper shows that, compared to the existing allocation policy at the time, the proposed policy can satisfy the same fairness criteria while resulting in an 8\% increase in aggregate quality-adjusted life years.   %

\cite{gupta2017maximizing} consider the problem of targeting individuals for a treatment or intervention. Clinical research studies often establish that a treatment has some sort of aggregate benefit in a large population. However, there may be significant heterogeneity in the treatment effect across different subgroups of that population. For example, an individual who is very sick may benefit less from a treatment than a less sick individual. A natural question, then, is how should one allocate the intervention to individuals in a population to maximize the aggregate benefit, subject to a limit on how many individuals may be treated (i.e., it is not possible to simply treat everyone), and a lack of knowledge of individual level treatment effects? The paper formulates this problem as a type of robust optimization problem, where the uncertainty set represents all heterogeneous treatment effects that are consistent with the aggregate characteristics of the research study. The paper uses data on a case management intervention for reducing emergency department utilization by adult Medicaid patients, and demonstrates settings under which the proposed robust optimization method provides a benefit over scoring rules typically used by medical practitioners. %

In a different direction, the paper of \cite{bertsimas2016analytics} considers the design of combination chemotherapy clinical trials for gastric cancer. The authors of the paper constructed a large database of gastric cancer clinical trials, and used this database to build two predictive models: one model to predict the median overall survival of patients in a trial based on the characteristics of the trial patients and the dosages of different drugs, and another model to predict whether the trial will exhibit an unacceptably high fraction of patients with severe toxicities. Using these predictive models, the paper then formulates a mixed-integer optimization model for deciding the next combination of drugs to test for given cohort of patients, so as to maximize the predicted median overall survival subject to limits on the predicted toxicity. The approach is evaluated using a method based on simulation and one based on matching a proposed clinical trial with the one most similar to it in the data; both evaluation methods suggest that the proposed approach can improve the overall efficacy of regimens that are chosen for testing in Phase III clinical trials.

\subsection{Hospital-level problems}
\label{subsec:healthcare_hospital}

At the hospital-level, the paper of \cite{rath2017integrated} studies the problem of staffing and scheduling surgeries at the Ronald Reagan Medical Center (RRMC) at the University of California, Los Angeles (UCLA), one of the largest medical centers in the United States. Patients who undergo surgery at the RRMC require three different types of resources: an anesthesiologist, a surgeon and an operating room. Prior to this research, anesthesiologists at RRMC would manually schedule surgeries, resulting in high operating costs. The paper presents a methodology, based on a two-stage robust mixed-integer optimization problem, that accounts for the uncertainty in surgery duration. The proposed method was implemented and is currently in use, saving RRMC approximately \$2.2 million per year through reduced anesthesiologist overtime. The follow-up paper of \cite{rath2018staff} focuses specifically on staff planning at the RRMC (deciding how many anesthesiologists are needed on regular duty and how many on overtime). While such decisions involve tangible, explicit costs (e.g., requiring an on-call anesthesiologist to come in requires a payment), they also involve implicit costs (e.g., having an anesthesiologist on call, but not calling them). The paper proposes an approach that learns such implicit costs from data on prior staffing decisions, and then uses these estimates within a two-stage integer stochastic dynamic program to make staffing decisions that leads to improvements in total (implicit plus explicit) cost. 

Another example of optimization being applied to hospital operations is the paper of \cite{zenteno2016systematic}. This paper describe the successful implementation by Massachusetts General Hospital (MGH) of a large-scale optimization model in order to reduce the surgical patient flow congestion in the perioperative environment, without requiring additional resources.
The paper studies an optimization model that rearranges the elective block schedule in order to smooth the average inpatient census by reducing the maximum average occupancy throughout the week. This model was implemented by MGH hospital and resulted in increasing the effective capacity of the surgical units at MGH. Outside of scheduling and staffing, the paper of \cite{bravo2019optimization} develops a linear optimization-based framework, inspired by network revenue management, for making resource allocation and case-mix decisions in a hospital network. Using data from an academic medical center, the paper shows how the approach can lead to better case mix decisions than current practice, which relies on prioritizing services based on traditional cost accounting methods.

Outside of optimization, the paper of \cite{ang2015accurate} focuses on predictive modeling, specifically the problem of predicting emergency department (ED) wait times in hospitals. Many approaches have been proposed to predict ED wait times, such as rolling average estimators, metrics based on fluid models and quantile regression. The paper proposes a method that combines the LASSO method of statistical learning \citep{tibshirani1996regression} with predictor variables that are derived from a generalized fluid model of the queue. %
The paper shows using real hospital data how the proposed method outperforms other approaches to predicting ED wait times, leading to reductions in mean squared error of up to 30\% relative to the rolling average method that is commonly used in hospitals today.

\subsection{Patient-level problems}
\label{subsec:healthcare_patient}

Finally, the increasing availability of electronic health record data has lead to the application of analytics methods to patient-level problems that are of a medical nature. For example, \cite{bastani2015online} consider a multi-armed bandit problem where at each period, the decision maker can choose one of finitely many decisions, and the reward of each decision is a function of a high-dimensional covariate vector. The problem is to find a policy that achieves minimal regret. To do this, the paper proposes a novel algorithm based on the LASSO method. The specific healthcare application studied in the paper is that of learning to dose optimally: a physician encounters patients described by various features (e.g., patient's gender, age, whether the patient has other pre-existing conditions, whether the patient is taking other drugs, and so on) and must prescribe a dose to each patient, but must learn the effect of different dosage levels on-the-fly. Using a dataset on warfarin dosing, the paper demonstrates how the approach can be effectively used to prescribe initial doses to patients who are starting warfarin therapy. 

Another example is \cite{bertsimas2017personalized}. This paper considers the problem of recommending a personalized drug treatment regimen to a patient with diabetes based on their medical attributes so as to minimize the patient's HbA1C (glycated hemoglobin A$_{1\mathrm{C}}$; a measure of a patient's baseline blood glucose level over the preceding 2-3 months). The paper proposes an approach that uses the $k$-nearest neighbors algorithm to determine, for a given patient, which patients are most similar to that patient in terms of their demographics (such as age and sex), medical history (such as days since first diabetes diagnosis) and treatment history (such as the number of regimens previously tried). By focusing on patients similar to the one at hand, one can obtain an estimate of what the patient's HbA1C will be under each possible drug regimen, and thereby recommend a good drug regimen. Using data from Boston Medical Center, the paper shows how the approach can lead to clinically significant reductions in HbA1C relative to the current standard of care.

\section{Conclusions and Future Directions}
\label{sec:conclusion}

In this paper, we have highlighted recent work in operations management, leveraging methodological advances in machine learning and optimization, that allows large-scale data to be used for complex decision-making. We conclude our review by discussing some methodological directions that we expect to grow in importance in the future.

\subsection{Causal inference} %

Machine learning is usually concerned with predicting a dependent variable using a collection of independent variables. In many OM applications, the decision enters into the machine learning model as an independent variable. For such settings, a more relevant perspective to take is in understanding the \emph{causal} impact of the decision on the dependent variable. This concern for the validity of decisions motivates the study of causal inference methods, which is a major area of focus in statistics and econometrics. 

The relevance of causal inference to empirical OM has been previously highlighted \citep{ho2017om}, but less research has highlighted the importance of considering causality in prescriptive modeling. We have already discussed one example, \cite{baardman2018detecting}, which develops a specialized estimation method for detecting customer-to-customer trends that uses the instrumental variables approach. Another example is the paper of \cite{bertsimas2016power}, which considers the problem of pricing from observational data. The paper shows that a standard approach to data-driven pricing, which involves fitting a regression model to price and then optimizing revenue as price multiplied by predicted demand, can actually be highly suboptimal when there exist confounding factors affecting both historical prices and demand. The paper develops a statistical hypothesis test for determining whether the price prescription from a predictive model is optimal, and apply this approach to an auto loan data set. 

Outside of revenue management, \cite{fisher2016value},  \cite{bandi2018opportunistic} and \cite{perakis2019} study causal inference and prescriptive modeling questions in supply chain management, particularly relating to the speed of delivery of online orders and subsequent product returns. \cite{fisher2016value} use a quasi-natural experiment to demonstrate the economic value of faster delivery for an online retailer.  \cite{perakis2019} take this one step further, by analyzing a transaction data set from a large fashion e-retailer and  establishing a causal relationship between product delivery gaps in the supply chain and product returns. Furthermore, the paper embeds this causal model within a data driven optimization framework in order to reduce product returns by optimizing product delivery services. Using the e-tailer's data, the paper shows that a 2-day delivery policy, that is, when all orders are attempted to be delivered within 2 days of the order placement, can lead to cost savings of as much as \$8.9 million every year through reduced product returns. %
\cite{bandi2018opportunistic} also investigate the causes of product returns, with a focus on how dynamic pricing can lead to more product returns in the online retail industry. In particular, using data from the same online retailer, they develop a logistic regression model that allows them to find that a post-purchase price drop is another cause of returns since it gives rise to opportunistic returns.

Recently, new approaches have also been emerging that combine machine learning with prescriptive analytics in order to deal with issues of heterogeneity. \cite{Stubhub} provide a framework for estimating price sensitivity when pricing tickets in a secondary market. Due to the heterogeneous nature of tickets, the unique market conditions at the time each ticket is listed, and the sparsity of available tickets, demand estimation needs to be done at the individual ticket level. The paper introduces a double-orthogonalized machine learning method for classification that isolates the causal effects of pricing on the outcome by removing the conditional effects of the ticket and market features. Furthermore, the paper shows how this price sensitivity estimation procedure can be embedded in an optimization model for selling tickets in a secondary market. This double orthogonalization procedure can also be applied to other settings, such as personalized pricing and estimating intervention effects in healthcare.

\subsection{Interpretability}

Another exciting future direction is in the development of interpretable models in operations management. In machine learning, an \emph{interpretable} model is one where a human being can easily see and understand how the model maps an observation into a prediction  \citep{freitas2014comprehensible}. Interpretability is a major area of research in machine learning for two reasons. First, interpretable models can provide insights into the prediction problem that black-box models, such as random forests and neural networks, cannot. Second, and more importantly, machine learning models often do not effect decisions directly, but instead make predictions or recommendations to a human decision maker; in many contexts (e.g., medicine), a decision maker is unlikely to accept a recommendation made by a machine learning model without the ability to understand how the recommendation was made. In addition, there is also growing legislation, such as the General Data Protection Rules (GDPR) in the European Union \citep{doshi2017towards}, requiring that algorithms affecting users must be able to provide an explanation of their decisions. 

While interpretable machine learning is still new to OM, some recent papers have considered interpretability in the context of dynamic decision making. \cite{bravo2018mining} develop an approach called MinOP (mining optimal policies), wherein one analyzes exact solutions of stochastic dynamic programs using interpretable machine learning methods; this approach provides a modeler with insight into the structure of the optimal policy for a given family of problems. \cite{ciocan2018interpretable} propose an algorithm for finding policies for optimal stopping problems in the form of a binary tree, similar to trees commonly used in machine learning; the paper shows how the proposed tree policies outperform (non-interpretable) policies derived from approximate dynamic programming in the problem of option pricing, while being significantly simpler to understand. With the continuing proliferation of machine learning-based decision support systems in modern business, we expect interpretable decision making to become a major area of research in OM.

\subsection{``Small data'' methods} 

One challenge that is being recognized with the increasing availability of data is that this data is often ``small''. This paradoxical statement refers to the fact that often, while the number of observations one might have access to is large, the number of parameters that one must estimate is also large. For example, imagine an online retailer offering a very large selection of products in a particular category; the retailer might be interested in estimating the demand rate of each product, but may only see a handful of purchases for each product in a year. The paper of \cite{gupta2017small} considers how to solve linear optimization problems in this ``small data'' regime and show that standard approaches in stochastic optimization, such as sample-average approximation, are suboptimal in this setting. The paper develops two different solution approaches, based on empirical Bayes methods and regularization, that enjoy desirable theoretical guarantees as the number of uncertain parameters grows, and showcases the benefits of these approaches in an online advertising portfolio application.

A different approach, which specifically considers prediction in the small data regime, can be found in the paper of \cite{farias2017learning}, which studies the following problem. A retailer offers a collection of products to a collection of users. For each product-user pair, we see whether the user has purchased that product or not, leading to a 0--1 matrix with users as rows and products as columns. Based on this data, we would like to estimate the probability of any given user purchasing any given product, leading to a matrix of user-product purchase probabilities; such a matrix is useful because it can guide product recommendations, for example. The problem is made challenging by its scale (as an example, Amazon has on the order of 100 million products and 100 million users) and the fact that there is a very small number of transactions per user, which inhibits one's ability to accurately estimate the user-product purchase probability matrix. To solve the problem, \cite{farias2017learning} leverage the availability of \emph{side information}. For example, Amazon may see other types of user-product interactions, e.g., when users view a product page, add a product to their wish list, add a product to their shopping cart, and so on. The data for each interaction can be represented as user-product matrix in the same way as the principal interaction (making a purchase); each matrix can thus be viewed as a slice of a three-dimensional tensor. The paper develops a novel algorithm for recovering slices of the data tensor based on singular value decomposition and linear regression; using data from Xiami, an online music streaming service in China, the paper shows how the proposed method can more accurately predict user-song interactions than existing methods that do not use side information. 

In an entirely different application, the tensor completion method of \cite{farias2017learning} has been used for the problem of designing a blood test for cancer in the paper of \cite{mahmoudi2018multi}. Specifically, an emerging idea for designing a blood test for cancer is to insert nanoparticles into a blood sample, and to measure the binding levels of these nanoparticles; this leads to data in the form of a three-dimensional tensor. While the raw data has too much noise to be directly used in a predictive model, \cite{mahmoudi2018multi} leverage the method of \cite{farias2017learning} to de-noise the data. Using data from a longitudinal study, \cite{mahmoudi2018multi} show how the resulting predictive models can predict whether a patient will develop cancer with diagnostically significant accuracy, 7-8 years \emph{after} the blood sample was drawn.

\subsection{New approaches to ``predict-then-optimize''}

Many of the papers described earlier rely on the ``predict-then-optimize'' paradigm: first build a predictive model using data, and then embed that model within the objective function of an optimization problem. Machine learning models are typically estimated so as to lead to good out-of-sample predictions; however, such models may not necessarily lead to good out-of-sample decisions. Rather than estimating models that minimize loss functions measuring predictive performance, the paper of \cite{elmachtoub2017smart} proposes minimizing a loss function that is related to the objective value of the decisions that ensue from the model. The paper shows how this new estimation approach leads to decisions that outperform the traditional approach in fundamental problems such as the shortest path, assignment and portfolio selection problems. Given the prescriptive focus of OM, this work highlights the potential opportunities for revisiting how machine learning models are constructed for prescriptive applications in OM.

\section*{Acknowledgments} 
The second author would like to acknowledge the support of NSF grant CMMI-1563343.

{\singlespacing
\bibliographystyle{plainnat}
\bibliography{ML_OM_literature.bib}

}
\end{document}